\documentclass[12pt,preprint]{aastex}
\usepackage{amsfonts,amsmath,graphicx,natbib}

\shortauthors{Bower et al.}
\shorttitle{Angular Size for GC Magnetar}
\begin{document}

\newcommand\degd{\ifmmode^{\circ}\!\!\!.\,\else$^{\circ}\!\!\!.\,$\fi}
\newcommand{\etal}{{\it et al.\ }}
\newcommand{\uv}{(u,v)}
\newcommand{\rdm}{{\rm\ rad\ m^{-2}}}
\newcommand{\msuny}{{\rm\ M_{\sun}\ y^{-1}}}
\newcommand{\mylesssim}{\stackrel{\scriptstyle <}{\scriptstyle \sim}}
\newcommand{\lsim}{\stackrel{\scriptstyle <}{\scriptstyle \sim}}
\newcommand{\gsim}{\stackrel{\scriptstyle >}{\scriptstyle \sim}}
\newcommand{\sci}{Science}

\def\kbar{{\mathchar'26\mkern-9mu k}}
\def\totd{{\mathrm{d}}}
\newcommand{\sgr}{SGR J1745-29}


\title{The Angular Broadening of the Galactic Center Pulsar \sgr:  A New Constraint on
the Scattering Medium}

\author{
 Geoffrey C.\ Bower\altaffilmark{1}, 
Adam Deller\altaffilmark{2}, Paul Demorest\altaffilmark{3},
Andreas Brunthaler\altaffilmark{4}, 
Ralph Eatough\altaffilmark{4},
Heino Falcke\altaffilmark{5,2,4},
Michael Kramer\altaffilmark{4} 
K.J. Lee\altaffilmark{4},
Laura Spitler\altaffilmark{4}
}

\altaffiltext{1}{Astronomy Department and Radio Astronomy Laboratory, University of California, Berkeley, 601 Campbell Hall \#3411, Berkeley, CA 94720, USA; gbower@astro.berkeley.edu}
\altaffiltext{2}{ASTRON, P.O. Box 2, 7990 AA Dwingeloo, The Netherlands}
\altaffiltext{3}{NRAO,520 Edgemont Road, Charlottesville, VA 22903-2475}
\altaffiltext{4}{Max-Planck-Institut f\"{u}r Radioastronomie, Auf dem H\"{u}gel 69, D-53121 Bonn, Germany}
\altaffiltext{5}{Department of Astrophysics, Institute for Mathematics, Astrophysics and Particle Physics (IMAPP), Radboud University, PO Box 9010, 6500 GL Nijmegen, The Netherlands}

\begin{abstract}
The pulsed radio emission from the Galactic Center (GC) magnetar \sgr\ probes the turbulent,
magnetized plasma of the GC hyperstrong scattering screen through both
angular and temporal broadening.  We present the first
measurements of the angular size of \sgr, obtained with the Very Long Baseline Array and the phased
Very Large Array at 8.7 and 15.4 GHz.  The source sizes 
are consistent with the scatter--broadened size of Sagittarius A* at each frequency,
demonstrating  that \sgr\ is also located behind the same hyperstrong scattering medium.  
Combining the angular broadening with temporal scattering obtained 
from pulsar observations provides a complete picture of the scattering properties.  
If the scattering occurs in a thin screen, then it must be at a distance
$\Delta \gsim 5$ kpc.  A best-fit solution for the distance of a thin screen
is $\Delta=5.9 \pm 0.3$ kpc, consistent with being located in the Scutum spiral arm.  
This is a substantial revision of the previously held model in which the scattering
screen is located very close to the GC.
As also discussed in \citet{Spitler},
these results suggest that GC searches can detect millisecond
pulsars gravitationally bound to Sgr A* with observations at $\gsim 10$ GHz and 
ordinary pulsars at even lower frequencies.
\end{abstract}


\section{Introduction}
\label{sec:intro}

A pulsar in orbit around Sagittarius A*, the black hole at the center of the Galaxy, has the
potential to provide a powerful probe of general relativity and the structure
of space-time in the strong field limit \citep[e.g.,][]{1986ARA&A..24..537B,1999ApJ...514..388W,2004ApJ...615..253P,2004NewAR..48.1413C,2012ApJ...747....1L}.  There have been numerous searches
for such a pulsar without success \citep[e.g.,][]{2000ASPC..202...37K,2006MNRAS.373L...6J,2009ApJ...702L.177D,2010ApJ...715..939M,2013IAUS..291...57S,2013IAUS..291..382E}.  
Searches of this kind
have been assumed to be challenging due to the presence of very large temporal broadening of
radio pulses due to strong interstellar scattering \citep{1997ApJ...475..557C} that is inferred
from the very large angular broadening seen for Sgr A* and other GC sources, with sizes 
of $\sim 0.5$ arcsec at 1.4 GHz \citep{1992ApJ...396..686V,2006ApJ...648L.127B}.  
The GC hyperstrong scattering screen has been modeled with
a turbulent, ionized thin screen at a distance
from the GC of $\Delta=133^{+200}_{-80}$  pc 
\citep{1998ApJ...505..715L}.
The large angular broadening and the proximity of the screen to the GC implies a temporal
broadening of pulsed emission of $\gsim 10$ seconds at 1.4 GHz,
eliminating any opportunity to detect pulsars at low frequencies.  Pulsar searching at higher
frequencies, however, is less sensitive due to the typical steep pulsar spectrum.
Detection of a pulsar in the GC provides us with the opportunity to simultaneously 
probe the angular and temporal properties of the scattering medium.

The pulsar \sgr\ was discovered serendipitously through its X-ray emission by the Swift
telescope \citep{2013ApJ...770L..24K}.  Subsequent NuSTAR observations detected periodic flux variations
with $P=3.76$ s and a hydrogen absorption column characteristic of a GC location
\citep{2013ApJ...770L..23M}. Chandra observations demonstrated that the source
was offset from Sgr A* by approximately 3 arcseconds, or a projected separation
of 0.1 pc \citep{chandrasgratel}.  The measured period
derivative implies a magnetic field $\sim 10^{14}$ G and a characteristic age of $9$ kyr.  
The spin down power is inadequate to account for the X-ray flux, identifying 
\sgr\ as a magnetar
\citep{1995MNRAS.275..255T}.

Radio pulsations were detected from \sgr\ at frequencies ranging from 1.4 to 20 GHz
\citep{2013arXiv1308.3147E,2013arXiv1305.3036S,Spitler}.  The radio observations confirm the X-ray pulse period and
period derivative.  They also determine the dispersion measure 
${\rm DM}=1778 \pm 3 {\rm\ pc\ cm^{-3}}$ 
and the rotation measure $RM=-66960 \pm 50 {\rm\ rad\ m^{-2}}$.
 These two measures of the line of sight plasma properties indicate
that \sgr\ is very likely to be physically close to Sgr A* and shares many of the same line of
sight characteristics; the DM is consistent with predictions of the galactic electron density
model  \citep{2002astro.ph..7156C}.

We present here the first very long baseline interferometric (VLBI) observations of \sgr\,
obtained with the Very Long Baseline Array (VLBA) and the phased array output of the Karl G. Jansky
Very Large Array (VLA).  These observations provide a direct measurement of the angular
size of \sgr.  Coupled with temporal broadening from pulsar observations \citep{Spitler}, we are able
to place a new set of constraints on the geometric model for hyperstrong scattering screen.
We present our observations and data analysis in \S~\ref{sec:obs} and discuss results in \S~\ref{sec:discussion}.

\section{Observations and Data Reduction}
\label{sec:obs}

\sgr\ has been observed 4 times using the VLBA + phased VLA as part of a campaign to 
measure the proper motion of \sgr\ (project codes BB336, BB337; 
Table~\ref{tab:obs}).
Each observation was 6 hours in duration and used Sgr A* as a delay and phase calibrator.  
In the first observation, the VLA was in the D configuration and the synthesized beam was large enough to encompass both 
\sgr\ and Sgr A* simultaneously.  For this observation only, an external phase reference calibrator
(J1752--3001) was observed for 1 minute every 5 minutes.  For all other observations, the
VLA synthesized beam was too small to include both Sgr A* and \sgr\ simultaneously, and so
we performed
a nodding cycle of total duration 2.5 minutes, with 36 seconds spent on Sgr A* and 96 seconds
on \sgr.

\begin{deluxetable}{cccc}
\tablecaption{Observations of \sgr}
\tablewidth{0pt}
\tablehead{
\colhead{Epoch} & \colhead{VLA configuration} & \colhead{phased VLA resolution}& \colhead{Observing frequency} 
\\
\colhead{(MJD)} &                             & \colhead{(\arcsec)} & \colhead{(GHz)} 
}
\startdata
56422 & D& 8.5  & 8.540 -- 8.796 \\
56444 & C&  2.6  & 8.540 -- 8.796 \\
56473 & C& 1.5   & 15.240 -- 15.496 \\
56486 & C&  2.6  & 8.540 -- 8.796 
\enddata
\label{tab:obs}
\end{deluxetable}

Due to scatter--broadening, Sgr A* appears as a Gassian with major axis size 
1.32 $\lambda^2$ mas at a wavelength $\lambda$ cm \citep{2006ApJ...648L.127B}.  
Nearby background sources are also scatter--broadened,
making it impossible to calibrate the longest VLBA baselines.  We were unable to 
calibrate the Saint Croix, Mauna Kea and Hancock stations; only short timeranges of useful data were obtained for Brewster, North Liberty and Owens Valley.  
The first observation which included an external calibrator provided better results for the 
more distant stations than the subsequent observations.
Difficulties with array phasing resulted in the loss of 2 hours of data from the second epoch, 
reducing sensitivity and $uv$ coverage.

Before VLBI correlation proceeded, the phased VLA data were analysed to provide contemporaneous timing 
information to enable pulsar gating \citep[e.g.][]{deller07a}.  For VLBI observations, the phased 
VLA records baseband voltage in VDIF format \citep{whitney09a},
which we then processed using DiFX correlator tools \citep{deller11a} and the DSPSR software
\citep{2011PASA...28....1V}.  
Then, for each observation of SGR~1745$-$29, a 256-channel coherently dedispersing
filterbank was applied to split the full band into 1-MHz wide channels,
and remove interstellar dispersion at the known DM of the pulsar.  This
was followed by full-Stokes detection and averaging into 4096-bin single
pulse profiles ($\sim$1 ms time resolution).  
By measuring the amplitude of the VLA 10 Hz switched noise calibration signal as a function of frequency 
and polarization, we were able to correct for differential gain and subtract the switched noise 
signal from the filterbank data.
The calibrated, 10-Hz-removed pulse profiles were
averaged into 5-minute integrations for MJD 56422 and 90-second
integrations for following epochs, from which we generated pulse times
of arrival (TOAs) in four 64-MHz subbands using the PSRCHIVE software 
\citep{2004PASA...21..302H,2012AR&T....9..237V}.
The pulse reference phase, average pulse period and DM for each observing session were
measured using the TEMPO software\footnote{http://tempo.sourceforge.net}.  This information was then
used to determine an on-pulse gate for VLBI correlation.

\begin{figure}[tb]
\includegraphics{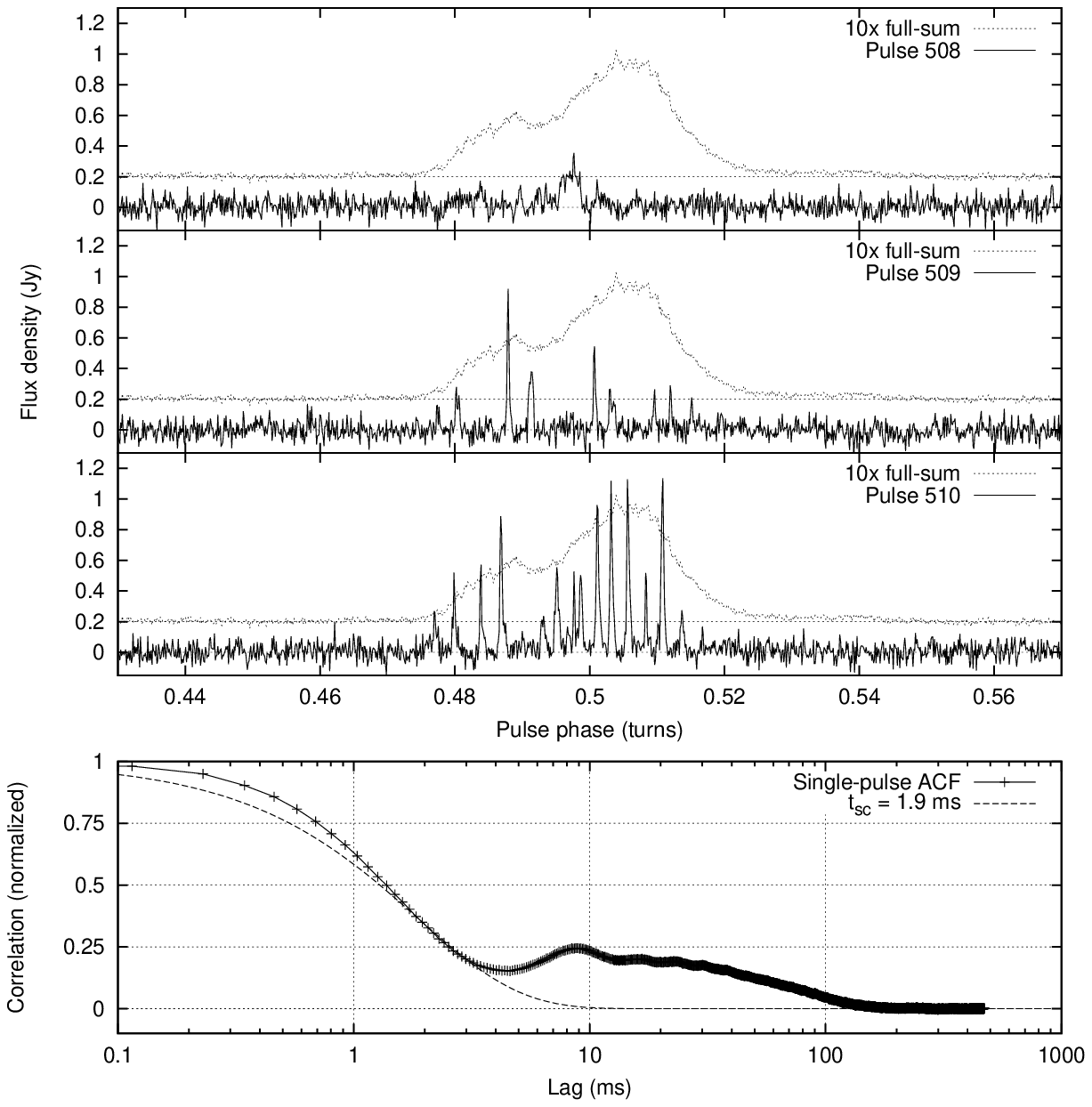}
\caption{Pulse shapes and temporal broadening.  The top three panels
show a series of three sequential individual pulses from phased VLA
observations on MJD 56486.  The average pulse profile, scaled by a
factor of 10 and offset by 0.2 Jy is also shown for reference.
Individual pulses show ms-timescale structure with amplitude $\sim$10
times the average profile.  These often contain quasi-periodically
spaced components with a typical separation of $\sim$10~ms (e.g. pulse
number 510 shown here).  The bottom panel shows the average single-pulse
autocorrelation function,  which has
a $1/e$ width of 1.9~ms.  The
expected ACF of intrinsically narrow single pulses convolved with an
exponentially decaying response of 1.9~ms timescale is shown for
comparison.
\label{fig:pulseshape}
}
\end{figure}

The average pulse profile (Figure~\ref{fig:pulseshape}) determined from our VLA
data consists of two broad features with an overall width $\sim$150~ms,
and a low-level trailing component extending another $\sim$100~ms.
However, individual pulses vary dramatically in shape, and exhibit
structure on ms timescales.  The width of these subpulses can be used to
constrain the temporal scatter broadening affecting the pulsar.  To
quantify this, we reprocessed our MJD 56486 data into 32768-bin single
pulse profiles (0.1~ms time resolution).  We computed the average
autocorrelation function (ACF) of all the high-resolution pulses using
an on-pulse window of 0.4$-$0.6~turns pulse phase.  To remove potential
bias due to any non-pulsar noise processes we subtracted the average ACF
measured in two adjacent off-pulse windows of identical length (pulse
phase 0.2$-$0.4 and 0.6$-$0.8).  The ACF shows a sharp, fully resolved component with a
$1/e$ width of 1.9~ms.  Temporal scatter-broadening in excess of this
value would push power to larger lags, so we interpret this as a
conservative upper limit on the scatter-broadening timescale at 8.7~GHz.
This upper limit is consistent with a measured timescale at 8.36 GHz of $0.3 \pm 0.4$ ms
\citep{Spitler}.

The VLBA and phased VLA data were correlated using the DiFX correlator \citep{deller11a}.  Three correlator passes were employed:
\begin{enumerate}
\item a gated pass using the ephemeris information derived from the VLA and the position of \sgr\ for all scans on \sgr\ or Sgr A* (the ``gated" pass);
\item an ungated pass using the position of \sgr\ for all scans on \sgr\ or Sgr A* (the ``ungated" pass); and
\item an ungated pass using the position of Sgr A* for all scans on \sgr\ or Sgr A* (the ``calibrator" pass).
\end{enumerate}
 A simple pulsar gate of width approximately 150 ms was used, boosting the S/N by a factor of $\sqrt{3.76/0.15}
 \sim 5$ on \sgr\ in the gated pass.
 
Data reduction used the AIPS software package \citep{2003ASSL..285..109G}
with the ParselTongue interface \citep{kettenis06a}.  Visibility amplitudes were
corrected for the logged system temperatures (and the nominal correction for the VLA, 
since no logged information was available). Delay calibration was performed using 
the ungated dataset with J1752--3001 in the first epoch and Sgr A* in subsequent epochs.  
Phase and amplitude self calibration with a timescale of 20s was performed using Sgr A*.  
The solutions and visibilities were flagged to remove outliers and Sgr A* was imaged using natural 
weighting with visibility weights raised to the power 0.5 to reduce the dominance of the VLA baselines.  
The calibration and flags were then applied to the gated and ungated correlator pass data.
 
Imaging at the position of \sgr\ requires an accurate subtraction of Sgr A*, which is only several hundred synthesized beamwidths away and around 5000 times brighter than \sgr\ in the ungated image.  
This is particularly problematic in the second and subsequent epochs, because Sgr A* moves through 
the sidelobe pattern of the phased VLA during observations of \sgr, suffering large gain variations. 
We remove Sgr A* by visibility peeling.
Phase and amplitude self--calibration solutions were determined for the VLA using 
the ungated pass visibilities and the Sgr A* model derived from the calibrator pass.
This calibration was applied to the gated pass dataset and the model of Sgr A* from the clean image 
was subtracted; then, the calibration was reverted using the AIPS task CLINV.  
In this way, optimal signal--to--noise was obtained for both the peeling solutions and the final gated image.

For the 3rd epoch at 15 GHz, Sgr A* is further into the sidelobes of the phased VLA, making the effective 
VLA gain in the direction of Sgr A* during the scans on \sgr\ even lower and more rapidly variable.  Unsurprisingly, this epoch suffered from much worse dynamic range issues resulting from the imperfect subtraction of Sgr A*.

\begin{figure}[tb]
\begin{tabular}{c}
\includegraphics[trim=0cm 1cm 1cm 1cm, clip, width=0.5\textwidth]{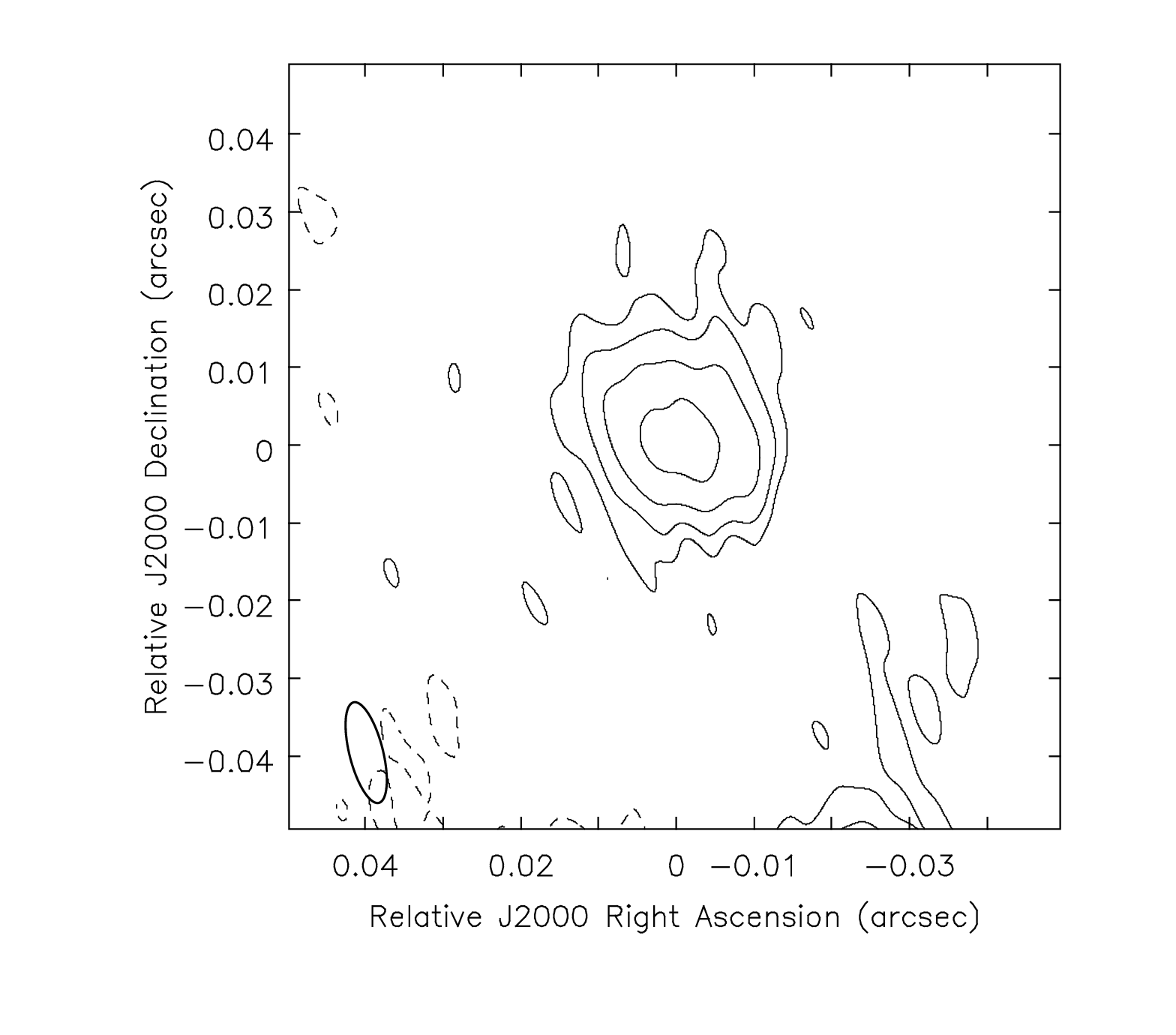} 
\includegraphics[trim=0cm 1cm 1cm 1cm, clip, width=0.5\textwidth]{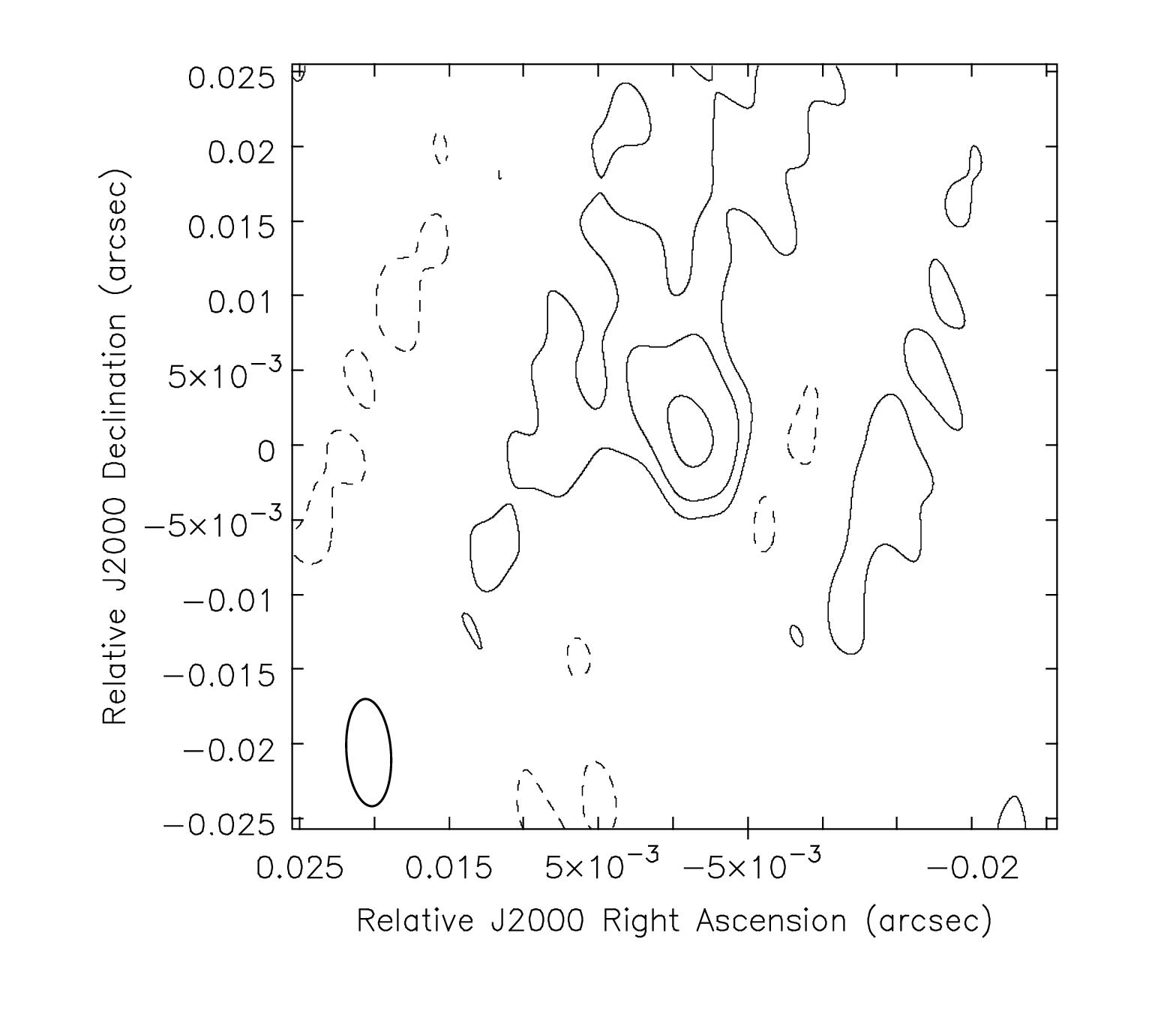}
\end{tabular}
\caption{Images of \sgr\ at 8.7 GHz (left panel, from the first epoch) and at 15.4 GHz (right panel, third epoch). Contours are shown at -2, 2, 4, 8, and (in the left panel only)16 $\sigma$.
The synthesized beam is shown in bold in the lower left corner of the image.
\label{fig:vlbiimages}
}
\end{figure}

After imaging with IMAGR using the same weighting scheme 
the sources were fitted using the AIPS task JMFIT.   The resulting  
deconvolved, scatter--broadened  sizes for \sgr\ and Sgr A* are listed in Table~\ref{tab:vlbifits}.
Images of \sgr\ at 8.7 and 15.4 GHz are shown in Figure~\ref{fig:vlbiimages}.  
The 15.4 GHz image clearly shows the residual effects of imperfect calibration and 
subtraction of Sgr A*, and highlight that the fitted values at this frequency should be
interpreted with caution.
 
\begin{deluxetable}{lccc}
\tablecaption{Scattered size fits from the VLBI observations}
\tabletypesize{\small}
\tablewidth{0pt}
\tablehead{
\colhead{Observation} & \colhead{Deconvolved \sgr\ fit} & \colhead{Deconvolved Sgr A* fit} & \colhead{VLBI beam} \\
\colhead{(MJD)} & \colhead{(mas)} & \colhead{(mas)} & \colhead{(mas)} 
}
\startdata
\multicolumn{4}{c}{8.7 GHz} \\
\hline
56422 & 16.1$^{+1.3}_{-1.4}$ $\times$ 9.4$^{+2}_{-2.3}$ at 87\degr$^{+12}_{-10}$
           & 16.48$^{+0.02}_{-0.02}$ $\times$ 7.71$^{+0.03}_{-0.03}$ at 83.3\degr$^{+0.1}_{-0.1}$
           & 13.3$\times$4.3 at 14\degr \\
56444 & 15.6$^{+2.3}_{-2.3}$ $\times$ 10.1$^{+3.3}_{-4.5}$ at 81\degr$^{+28}_{-20}$ 
           & 15.9$^{+0.1}_{-0.1}$ $\times$ 6.8$^{+0.2}_{-0.3}$ at 89.5\degr$^{+0.6}_{-0.6}$
           & 17.6$\times$8.1 at 9\degr \\
56486 &  16.1$^{+3.2}_{-2.8}$ $\times$ 10.9 $^{+4.1}_{-5.4}$ at 71\degr$^{+23}_{-10}$
           &  16.07$^{+0.01}_{-0.02}$ $\times$ 8.02 $^{+0.02}_{-0.02}$ at 82.0\degr$^{+0.1}_{-0.1}$
           & 15.0$\times$8.0 at 6\degr \\
Average & $16.0^{+1.1}_{-1.1} \times 9.7^{+1.6}_{-1.9}$ at 82\degr$^{+10}_{-7}$
           & $16.21^{+0.01}_{-0.01} \times 7.92^{+0.02}_{-0.02}$ at 82.7\degr$^{0.1}_{-0.1}$
           & \dots \\
\hline
\multicolumn{4}{c}{15.4 GHz}\\
\hline
56473 & 6.8$^{+2.1}_{-2.6}$ $\times$ 4.4$^{+1.4}_{-4.4}$ at 35\degr$^{+42}_{-30}$   
           &  5.44$^{+0.00}_{-0.01}$ $\times$ 3.66$^{+0.01}_{-0.00}$ at 81.5\degr$^{+0.1}_{-0.1}$ 
           &  7.3$\times$3.0 at $-$5\degr \\
\enddata
\label{tab:vlbifits}
\end{deluxetable}

\section{Discussion and Conclusions}
\label{sec:discussion}

The measured size for Sgr A* is consistent with past measurements of the $\lambda^2$ scattering law.
\citet{2004Sci...304..704} obtained a size of $17.6 \pm 0.3 \times 8.5^{+1.3}_{-1.9}$ mas in position angle
$82 \pm 4$ degrees at a frequency of 8.4 GHz.  Scaling by the $\lambda^2$ law to 8.7 GHz, the
observation frequency of these new measurements, we find a size $16.4 \pm 0.3 \times 7.9^{+1.2}_{-1.8}$ mas, which is in 
agreement with the measured sizes for Sgr A* in Table~\ref{tab:vlbifits}.
\sgr\ has an angular size that is statistically indistinguishable from that of Sgr A* at 8.7 GHz.  
The small difference in the 15.4 GHz fits
is almost certainly due to residual calibration/subtraction errors.
The similarity between the angular broadening seen for \sgr\ and Sgr A* confirms that \sgr\ must be 
close to the GC and share a similar line of sight scattering medium.  In particular, it rules out the 
possibility that \sgr\ resides in front of the hyperstrong GC scattering screen, which would otherwise
be a possible explanation for the low temporal broadening discussed below.

The temporal scattering, $\tau$, is related to the angular broadening due to 
the longer path length required for scattered photons to reach the observer.  The 
time delay is proportional to the path integral of the solid angle and, 
therefore, has a different
weighting with distance than the angular broadening angle.  For the same angular broadening, the temporal 
broadening is much larger if the scattering occurs close to the source.
The angular and temporal scattering scales for a thin scattering screen are related by the following equation:
\begin{equation}
\tau = 6.3 {\rm\ s\ } \times {\left( D \over 8.5 {\rm kpc} \right)} {\left( \theta_1 \over 1.3 {\rm\ arcsec} \right)^2} \left( {D \over \Delta} - 1 \right) \nu^{-4},
\label{eqn:tau}
\end{equation}
where $D = 8.3 \pm 0.3$ kpc \citep{2010RvMP...82.3121G}
is the distance to the GC, $\Delta$ is the distance from the GC to the pulsar, 
and $\nu$ is the observing frequency
in GHz \citep{1997ApJ...475..557C}.  $\theta_1$ is the observed angular size 
extrapolated to a frequency of 1.0 GHz using a scaling of $\nu^{-2}$.  This 
equation assumes a symmetric scattering
medium, which is accurate to a factor of 2 in this case.  The effects of anisotropy on scattering tend to
be small \citep{2002ApJ...576..176C}.  We use the geometric mean of the major and minor scattering
axes at 8.7 GHz for our calculations, $\theta=12.5 \pm 1.2$ mas, implying $\theta_1=945 \pm 91$ mas.

\citet{1998ApJ...505..715L} determined $\Delta=0.133^{+0.2}_{-0.08}$ kpc, implying that 
$\tau$ ranges from 18 to 120 s at 1.4 GHz
for the $1\sigma$ range of $\Delta$. The maximal broadening occurs for the screen closest to Sgr A*.  
Thus, this estimate for $\Delta$ is clearly inconsistent with the detection of 
pulsed emission at frequencies below a few GHz in a pulsar with even a period as large as $P=3.7$ s.

\begin{figure}[tb]
\includegraphics[width=\textwidth]{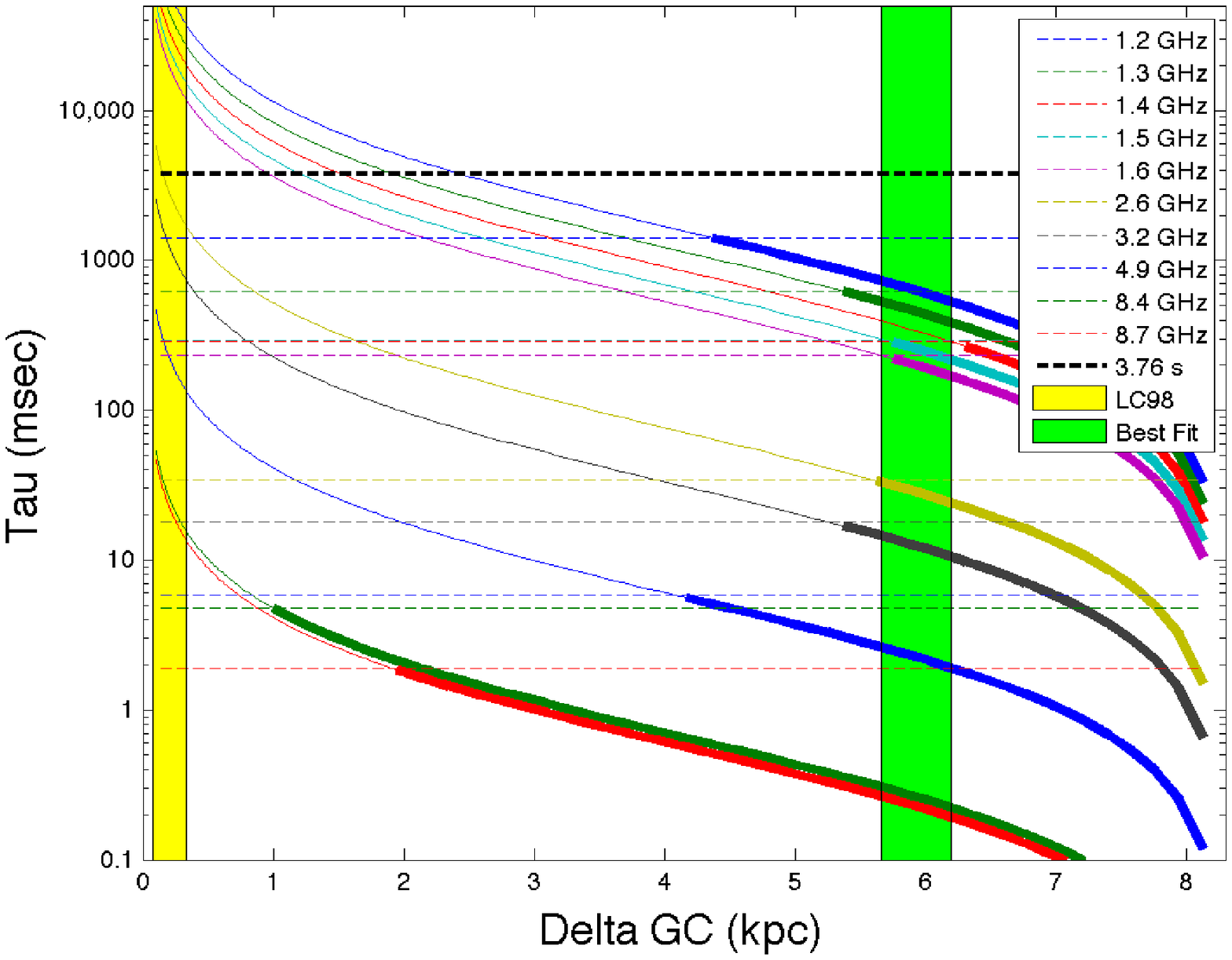}
\caption{Temporal broadening of pulsed radio emission as a function of distance $\Delta$ from the 
GC.  Horizontal dashed lines indicate the measured constraints 
from \citet{Spitler},
averaged in 100 MHz bins, and our simultaneous 8.7 GHz VLA observations.
Solid lines
are theoretical curves based on Equation~\ref{eqn:tau}.  Permitted parameter space is at values of $\Delta$
where $\tau$ is less than the observed values; these are highlighted as thick solid
lines.  
The dashed black line indicates the pulse period of 3.76 s.
The solid yellow region is the solution for $\Delta$ from \citet{1998ApJ...505..715L}.  
The solid green region is the best-fit solution $\Delta=5.9 \pm 0.3$ kpc from this paper.
\label{fig:delta}
}
\end{figure}

We combine our angular broadening measurement with 
temporal broadening measurements from \citet{Spitler} and our 8.7 GHz VLA data
to constrain $\Delta$ (Figure~\ref{fig:delta}).  Solutions are permitted for values
of $\Delta$ where $\tau$ falls below the measured pulse width.  We consider only measurements
at frequencies $< 10$ GHz; above this frequency, the scattering timescale becomes much smaller than the
intrinsic pulse widths. 
A thin screen solution is consistent for all frequencies with $\Delta \gsim 5$ kpc.  
A best-fit solution using a scattering time scale at 1 GHz
of $1.3 \pm 0.2$ s \citep{Spitler} for a thin screen scattering distance is $\Delta=5.9 \pm 0.3$ kpc.

This estimate for $\Delta$ is significantly larger than the upper bound on $\Delta$ set by 
the OH/IR star distribution and the number counts of extragalactic background sources
\citet{1998ApJ...505..715L}.  Recent results provide a counter to each of these arguments.  First, 
while a subset of the OH/IR stars within the
central 0.5 degree do show very strong scattering, scattering angles that are smaller by only a factor of two
are seen as far away as 6 deg from Sgr A*.  The discovery of patchiness in the scattering medium
\citep{2013arXiv1306.1842R} suggests that the line of sight towards
Sgr A* may not be as unique as previously stated and that the assumption of a single uniform screen for Sgr A*
and the OH/IR stars is not sufficient.
Second, new extragalactic background source counts indicate that deficit in source counts previously found
may not be as significant as previously estimated.
\citet{1998ApJ...505..715L} reported
a significant deficit in  1.3 and 1.7 GHz surveys of the GC, which was accounted for by very strong
scattering that leads to angular broadening of $\gsim 100\arcsec$ for extragalactic sources.  
However, \citet{2013arXiv1306.1842R} recently used
154 and 255 MHz extragalactic source counts to show that only a small fraction of sources are missing due to
scattering.  Additionally, other scattering sources that are either very close to or more distant from
the GC could affect extragalactic background sources without influencing Sgr A* or \sgr.

Additionally, the detection of Sgr A* at 330 MHz 
demonstrates that the free-free optical
depth towards Sgr A* is $\gsim 1$ \citep{2004ApJ...601L..51N}.
   As \citet{1992ApJ...396..686V} demonstrate, a low optical depth requires that the scattering 
screen be located at a distance $\Delta > 0.1 D \sim 0.8 $ kpc, consistent with our much larger
distance.  

Since the hyperstrong scattering medium
is not local to the GC,  it must be located in one or more scattering screens along the line of sight to the 
GC.  A few other lines of sight, such as towards NGC 6334B and the Cygnus region, have comparable
or larger scattering measures, which are argued to be due to random superposition of very turbulent clouds 
\citep{1994MNRAS.269...67W,1998ApJ...493..666T,2001ApJS..133..395D}.  
The characteristic turbulent clump size in the interstellar medium of 50 pc \citep{1991ApJ...372..132F}  
has a sufficiently large angular size at 5.9 kpc to be responsible for Sgr A*, \sgr, and many of
the OH/IR masers.
A distance of 5.9 kpc from the GC falls within or close to the Scutum arm, which hosts giant molecular
clouds and HII regions that could be responsible for the scattering \citep{2009ApJ...700..137R}.

Radio continuum measurements towards the GC
constrain the physical conditions of the scattering screen with a
relationship between $\Delta$,
$l_0$, the outer scale of the turbulent electron density (in pc), and $T_e$, the
electron temperature
 \citep[in K;][]{1998ApJ...505..715L}:
\begin{equation}
\log{l_0^{2/3} T_e^{-1/2}} \approx -7.0 + 2 \log{\Delta \over {\rm 150\, pc}}.
\end{equation}
For $\Delta=5.9$ kpc, we find solutions at typical warm and hot ionized medium temperatures ($T_e = 10^{4}$  and $10^{6}$ K)
of $l_0 = 10^{-3}$ pc  and $l_0 = 3 \times 10^{-1}$ pc.  These estimates for the outer scale are consistent 
with the lower limit of $\sim 3 \times 10^{-2}$ pc found from other galactic measurements 
\citep{1995ApJ...443..209A}.  Thus, the scattering screen
appears to have properties that are consistent with other turbulent regions in the interstellar medium.

The scatter broadening may occur in a thick screen or other unconventional geometric model
in which Equation~\ref{eqn:tau} does not hold.  \citet{Spitler} discuss thick
screen models and the evidence that may support that interpretation.
Our results demonstrate, however, that the angular broadening of \sgr\ resembles
that of Sgr A* and we have very strong constraints on that image:  the angular size scales as $\lambda^2$
and the angular size at a given wavelength is very well parametrized as a Gaussian \citep{2004Sci...304..704}.
These facts point to a conventional thin screen model for the pulsar scattering.

The absence of strong temporal broadening for the \sgr, coupled with clear evidence that \sgr\ sees
the same scattering medium as Sgr A* via the angular broadening, 
makes clear that millisecond pulsar companions to Sgr A* can be detected at 
frequencies $\gsim 10$ GHz, while ordinary pulsars can be detected at frequencies of a few GHz.  
Searches at 10 GHz will face temporal broadening of less than 
a millisecond, making observations sensitive to even the fastest spinning pulsars orbiting the black hole.
It is surprising that past attempts to search for GC pulsars have not been successful given
the expected large number of compact objects near Sgr A* and the broad frequency range
of those surveys \citep[][and references therein]{2010ApJ...715..939M}.
The detection of a pulsar through the hyperstrong scattering medium
demonstrates that past failures to discover a pulsar orbiting Sgr A*
must inform us about search methodologies and/or the population 
of GC pulsars.

\acknowledgements
The National Radio Astronomy Observatory is a facility of the National Science Foundation operated under cooperative agreement by Associated Universities, Inc. We thank H.J. van Langevelde for useful comments.


\end{document}